\documentclass[%
 aip,
 amsmath,amssymb,
 reprint,%
]{revtex4-1}

\usepackage{graphicx}
\usepackage{dcolumn}
\usepackage{bm}

\usepackage[utf8]{inputenc}
\usepackage[T1]{fontenc}
\usepackage{mathptmx}
\usepackage{etoolbox}

\makeatletter
\def\@email#1#2{%
 \endgroup
 \patchcmd{\titleblock@produce}
  {\frontmatter@RRAPformat}
  {\frontmatter@RRAPformat{\produce@RRAP{*#1\href{mailto:#2}{#2}}}\frontmatter@RRAPformat}
  {}{}
}%
\makeatother
\begin{document}

\preprint{AIP/123-QED}

\title[Chaotic VIR's on a elliptical cylinder]{Chaotic vortex induced rotation of an elliptical cylinder.}
\author{F. Mandujano}%
\email{frmas@ciencias.unam.mx}
\author{E. V\'azquez-Luis}%
 \email{edgar.vazquez@ciencias.unam.mx}
\affiliation{ 
Laboratory of Fluids, Physics Department, Faculty of Science, National University of Mexico
}%


\date{\today}

\begin{abstract}
Non-linear oscillations of an elliptical cylinder, that can rotate about an axis that passes through its symmetry axle due to a torsional spring and hydrodynamic torque produced by the flow of a Newtonian fluid, were analysed in terms of a single parameter that compares vortex shedding frequency with the torsional spring's natural frequency. The governing equations for the flow coupled with a rigid body with one degree of freedom, were solved numerically using the Lattice Boltzmann Method (LBM). The Reynolds number used was $Re=200$ which, in absence of torsional spring, produces chaotic oscillations of the elliptical cylinder. When the torsional spring is included, we identified three branches separated by transition regions when stiffness of the restorative torque change, as in the case of vortex induced vibrations (VIV's). However in this case, several regions presenting chaotic dynamics were identified. Two regions with stable limit cycles were found when both torques synchronized and when stiffness of the torsional spring is big enough so that the ellipse's oscillation is small. 
\end{abstract}

\maketitle

\begin{quotation}
Dynamics of an oscillating elliptical cylinder due to flow of a viscous fluid and torsional spring is presented. The Reynolds number is fixed at $Re=200$ and spring stiffness is modified. We identified three regions: excitation, lock-in and desynchronization regions as in the case of VIV's on circular cylinders. The problem has three fixed-points whose properties depends on the ellipse's oscillations degree of confinement, which can be changed by varying the spring's stiffness.     
\end{quotation}

\section{Introduction}
The study of fluid-solid interactions has been a problem of great interest for a long time from varying points of views. Vortex induced vibrations on solid structures as elastically mounted cylinders~\cite{williamson,sarpkaya,bearman1984}, ellipses~\cite{bhattacharya16,abu22,NAVROSE201455}, flexible cylinders and cables~\cite{newman97,brika,yamamoto04} are subjects of research in many fields of engineering due to their wide range of practical applications. In addition, from a mathematical point of view, the study of non-linear phenomena due to flow of a viscous fluid around rigid bodies when different numbers of degrees of freedom are considered~\cite{jauvtis03,reyes,dahl06} and chaotic dynamics of the rigid body due to the flow~\cite{weymouth,rosen2017,zhao14}, has attracted the attention  of several researchers because of modern development in computers and numerical methods. 

Movement of an elastically mounted cylinder is modeled using linear and non-linear forced oscillators, depending on the value of the so called reduced velocity~\cite{sarpkaya,williamson}; a parameter that compares vortex shedding frequency and the cylinder's oscillation frequency without flow. In such case when only translational degrees of freedom are considered, the cylinder's movement is modeled as a forced-damped oscillator in the lock-in region, where amplitudes of oscillation transversal to the flow direction are the largest~\cite{sarpkaya,williamson}. 

When rotational degrees of freedom are considered, movement of the rigid body is due to an auto-rotation effect that was originally studied by J. C. Maxwell in the 1890's~\cite{maxwell} and was later characterized by H. J. Lugt~\cite{lugt,lugt83}. Vortex shedding behind the body produces a torque that oscillates with time and, depending on the Reynolds number, behaves as a non-linear function of angular position, angular velocity and time. Non-linear dynamics observed in 2D and 3D have been analyzed from different points of view; for instance in the design of power extraction systems~\cite{bhattacharya16} and the study of non-linear oscillators~\cite{weymouth,rosen2017}. 

In this work we present numerical solutions of the flow of a viscous fluid around an elliptical cylinder that can rotate due to external restorative torque, with stiffness $k$, and hydrodynamic torque due to the flow. Reynolds number was fixed at $Re=200$ where oscillations of an elliptical cylinder without restorative torque are known to be non-linear~\cite{bhattacharya16,weymouth}. 

A detailed analysis of cylinder and flow dynamics was made as function of a single parameter, proportional to the natural frequency of restorative torque. The problem shows a similar behavior to that observed in cylinders with translational degrees of freedom: an excitation region where oscillation is periodic and a lock-in region where amplitude of oscillation reaches maximum values. A third region can be identified with a decoherence region, where amplitude of oscillation is diminished. 

These regions are separated by transition regions and both show different non-linear behavior and chaos. The transition between decoherence and lock-in regions is a chaotic region with an associated strange attractor. In the transition between excitation and lock-in regions, forced-damped periodic oscillations were observed when $k\rightarrow \infty$. Then, as $k$ is diminished, the cylinder's oscillations evolve into a limit cycle  through a region where the ellipse's amplitude in each oscillation changes erratically. Vorticity generated in the flow depends on the elliptical cylinder's degree of confinement which is controlled with the spring's stiffness.   

This article is organized with the problem statement in section \ref{ps} including a brief description of the numerical method. In section \ref{res} the most relevant results are presented and, in section \ref{dis}, we present a discussion of our main findings. 

\section{Problem statement}
\label{ps}

\begin{figure}[ht] 
        \centering \includegraphics[width=1.2\columnwidth]{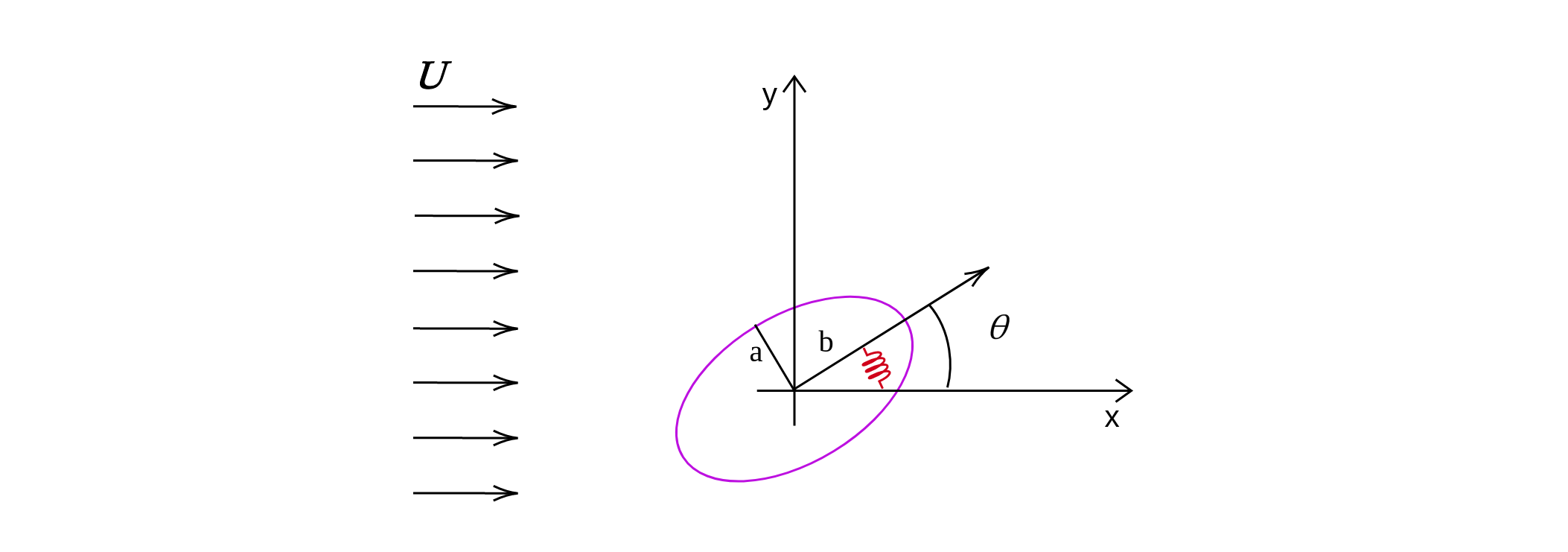}        
        \caption{\label{fig:elipse} Diagram of the problem. The flow around an elliptical cylinder that can rotate around its symmetry axis due to hydrodynamic and restorative torques. The flow is considered to be uniform, with velocity ${\bf U}$ far form the body, $a$ and $b$ are the minor and major axis of the ellipse respectively and $\theta=\theta(t)$ is the angle of rotation.}
 \end{figure}
The problem consists on uniform flow of a Newtonian viscous fluid, with mass density $\rho$ and kinematic viscosity $\nu$, around an elliptical cylinder that can rotate around its symmetry axis due to hydrodynamic torque and is attached to a torsional spring with constant stiffness $k$. The elliptical cylinder has a mass density $\rho_{s}$, $a$ and $b$ are the semi-minor and semi-major axes respectively and $\theta$ is angle of rotation, as shown in Figure \ref{fig:elipse}. We did not include structural damping to enforce the ellipse's oscillations in this work. The torsional spring's equilibrium angle is fixed at $\theta_e=0$.

In order to work with non dimensional variables, space is scaled with $b$, velocities with velocity far from the body $U$, and times with $b/U$. With this choice, the problem is characterized by four non dimensional parameters: Reynolds number $Re=\frac{U b}{\nu}$, modified Strouhal number $S_t=\sqrt{\frac{k}{I}}\frac{b}{U}$ that compares natural frequency of restorative torque with a scale of the vortex shedding frequency, ratio between semi-major and semi-minor axis $\alpha=\frac{b}{a}$, ratio between solid and fluid mass densities $m^* = \frac{\rho_{s}}{\rho_{f}}$. 

Considering $\mathbf{u}(\mathbf{r} , t )$ and $P(\mathbf{r} , t )$  as non dimensional velocity and pressure fields (pressure is scaled with characteristic viscous pressure  $\frac{\rho_f \nu U}{b}$), the governing equations for this problem are given by
\begin{eqnarray}
 \nabla\cdot\mathbf{u}&=&0,\\
Re\left(\frac{\partial \mathbf{u}}{\partial t} + \mathbf{u} \cdot \nabla \mathbf{u}\right) &=& -\nabla P + \nabla^2 \mathbf{u},
\end{eqnarray}
which correspond to non dimensional Navier-Stokes equations for an incompresible flow. Boundary conditions in this case are
\begin{eqnarray}
 \mathbf{u}(\mathbf{r}_s,t)&=&\mathbf{v_s}(\mathbf{r}_s,t),\label{bc1} \\
 \mathbf{u}(|{\bf r}| \rightarrow \infty ,t)&=& \hat{i},\label{bc2}
 \end{eqnarray}
where $\hat{i}$ is the unitary vector in the x direction, $\mathbf{r}_s$ is a point in the ellipse's surface and $\mathbf{v}_s(\mathbf{r}_s,t)$ its velocity. The above conditions correspond to non-slip at the ellipse's surface and a uniform flow far from the body.
Rotation of the ellipse is solved using the Newton equation for torques given by
\begin{equation}
\frac{d\mathbf{\omega (t)}}{dt}=\frac{\tau}{I^*} - S_t^2 (\theta-\theta_e),\label{nec}
 \end{equation}
where $I^*=\pi m^* \alpha (1+ \alpha^2)/4 $ is a reduced moment of inertia and dimensionless hydrodynamic torque is
\begin{equation}
\tau=\frac{1}{\rho_f U^2 a^2} \int_{S} \hat{{\bf k}} \cdot (r_s -\mathbf{R}) \times \boldsymbol{\sigma} \cdot \hat{n} dS.
\end{equation}
where $\mathbf{R}$ and $\omega(t)$ are the position of the ellipse's center of mass and angular velocity of rotation, respectively; $\boldsymbol{\sigma}$ represents the stress tensor of a Newtonian fluid.

\subsection{Numerical Method}
\label{nm}
To find numerical solutions for the above problem we used a two dimensional, nine neighbors (D2Q9) lattice-Boltzmann model~\cite{He98}. The proposed algorithm, that includes moving immersed boundaries, has been validated in previous works~\cite{reyes,mandujano18}. 

In this method space is discretized using a squared lattice. Lattice spacing as well as time steps can be 
conveniently set to unity. The fluid's state  at the node with vector position $\mathbf{r}$ at time $t$, is described by the particle distribution function $f_k(\mathbf{r},t)$ that evolves in time and space according to 
\begin{equation}
 f_k(\mathbf{r}+\mathbf{e}_k,t+1) = 
  f_k(\mathbf{r},t) -\dfrac{1}{\tau}\left[f_k(\mathbf{r},t)- f_{k}^{(eq)}(\mathbf{r},t)\right],\label{eq:lbed}
\end{equation}
where $\tau$ is a relaxation time related to the fluid's kinematic viscosity $\nu = (\tau- 1/2)/ 3$. Equilibrium distribution function $f_{k}^{(eq)}$ is given by
\begin{equation}
 f_{k}^{(eq)}(\mathbf{r},t) = w_k\rho \left( 1 + 3\mathbf{e}_k\cdot\mathbf{u} + \dfrac{9}{2}(\mathbf{e}_k\cdot\mathbf{u})^{2}-\frac{3}{2}\mathbf{u}^2\right),\label{eq:feq}
\end{equation}
which corresponds to a discrete Maxwell distribution function for thermal equilibrium. In the above expressions, macroscopic density and velocity fields are computed using
\[
 \rho(\mathbf{r},t) = \sum_{k}f_k(\mathbf{r},t) \quad \textmd{and} \quad
 \rho {\mathbf{u}}(\mathbf{r},t) = \sum_{k}\mathbf{e}_k f_k(\mathbf{r},t),
\]
and microscopic velocities $\mathbf{e}_k$ are given by
\begin{eqnarray}
 \mathbf{e}_0 &=& (0,0),\nonumber \\
 \mathbf{e}_k &= & \left(\cos(\pi(k-1)/2),\sin(\pi(k-1)/2)\right), \nonumber\\ && \textmd{for} \ \ k=1,\dots,4,\nonumber \\
 \mathbf{e}_k &=& \sqrt{2}\left(\cos(\pi(k-9/2)/2),\sin(\pi(k-9/2)/2)\right),\nonumber \\ &&\textmd{for} \ \  k=5,\dots ,8,\nonumber
\end{eqnarray}
where $w_0=4/9$, $w_k=1/9$ for $k=1,\dots,4$ and $w_k=1/36$ for $k=5,\dots,8$. Notice that, with this set of microscopic velocities, expression (\ref{eq:lbed}) is always evaluated at lattice points. It is well known that the above procedure approximates solutions to the Navier-Stokes equations in the limit of small Mach numbers \cite{He98}.

Equation (\ref{eq:lbed}) provides an explicit algorithm for updating all distribution functions $f_k$ at a given node in the lattice, as long as its 8 nearest neighbouring nodes are inside the fluid's domain. For nodes adjacent to a solid wall, distribution functions coming from neighbouring nodes outside the fluid domain must be provided. We chose the set of boundary conditions proposed by Guo and Zheng in Ref.~\cite{guo02a} for curved rigid walls, which approximately gives the no-slip boundary condition (\ref{bc1}). The force and torque acting on the body were computed using the momentum-exchange method of Mei et. al. in Ref.~\cite{mei02} and then equation (\ref{nec}) is solved using a leap-frog method.

In order to simulate an infinite domain, a constant velocity was enforced at the inlet whilst in the rest of the boundaries a stress-free condition was implemented (see for details F. Mandujano and C. M\'alaga~\cite{mandujano08}). The domain's size was chosen to be big enough such that effects due to lateral walls were minimized but the internal boundary could have enough number of points. The numerical scheme was implemented to run in parallel in Graphical Processor Units (GPU) due to the large number of nodes involved in these simulations. 

\section{Results}
\label{res}
Simulations started with a uniform flow and the ellipse fixed in space, until the fluid reached a time dependent flow with vortex shedding. Then the solid, initially at $\theta(0)=\theta_0$, was allowed to rotate following equation (\ref{nec}). Aspect ratio was fixed at $\alpha=2$ with $m^*=1$ and, as mentioned before, the Reynolds number was fixed at $Re=200$. Numerical experiments were performed varying the modified Strouhal number $S_t$. 
\begin{figure}[ht]
\centering
  \includegraphics[width=0.8\linewidth]{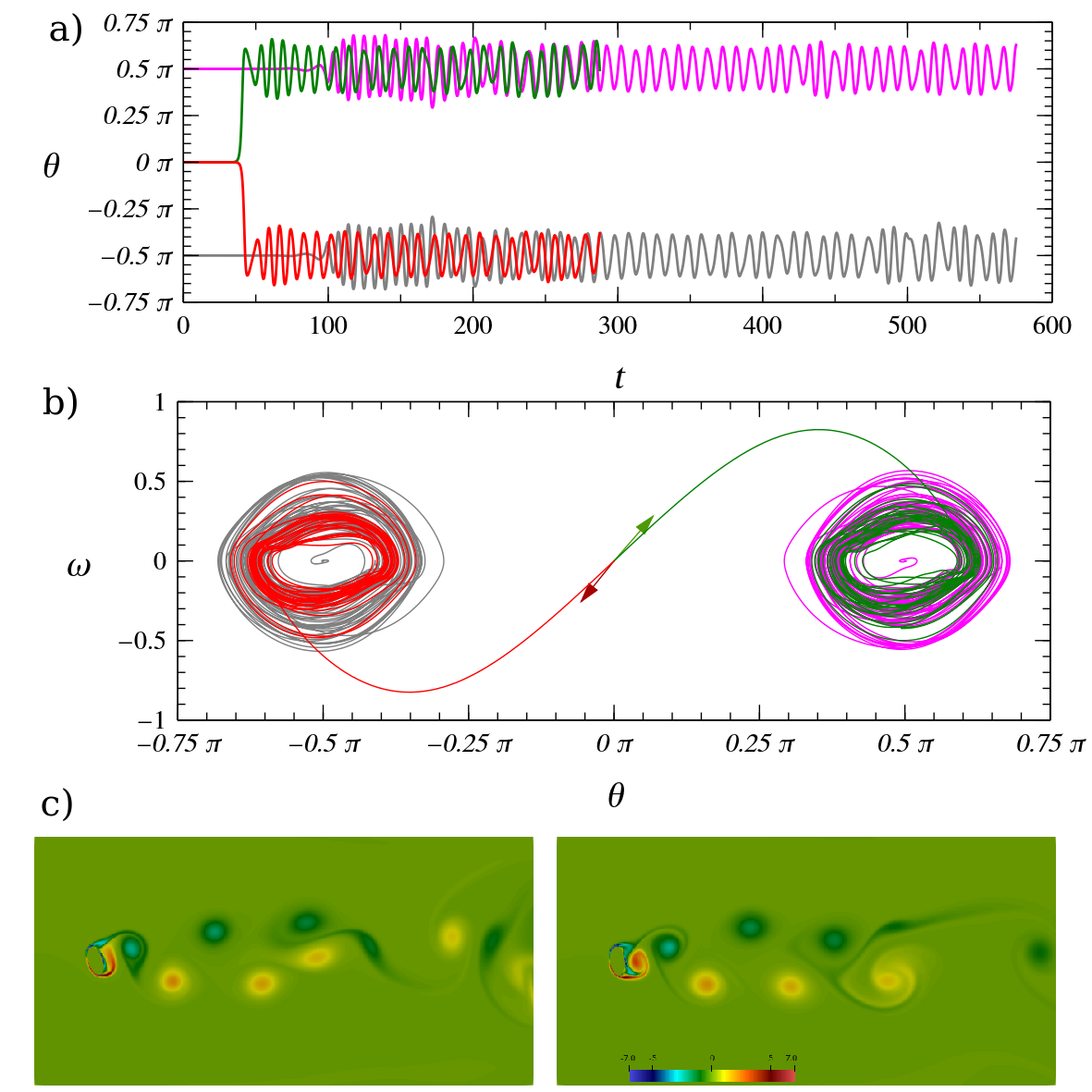}
  \caption{a) Rotational angle as function of time for different initial conditions with $S_t=0$. b) Phase space trajectories correspond to those shown in (a), arrows indicate the trajectories' starting point. c) Vorticity field at two different times separated approximately by one cycle.}
  \label{fig:pstr0}
\end{figure}%

In absence of restorative torque ($S_t=0$), the hydrodynamic torque oscillates with time in a non-periodic fashion with a non-zero mean. Hence, for any choice of initial condition $\theta_0 $ (initial angular velocity is set to zero in all cases), the ellipse's semi-mayor axis will oscillate around the vertical axis, as shown in Fig. \ref{fig:pstr0}-(a). When $\theta_0$ is chosen in the neighborhood of $\theta_0 = 0$, trajectories in phase space are expelled from the origin and, as time grows, are attracted making open orbits around either $\theta_c = \pi/2$ or $\theta_c = -\pi/2$ (see fig \ref{fig:pstr0}-(a)). The elliptical cylinder is then in an unstable equilibrium position and starts to move due to auto-rotation~\cite{lugt,weymouth2014}. We identified $\theta_s = 0$ as a fixed-point that behaves as a saddle point (see Fig. \ref{fig:pstr0}-(b)).  

Similarly, when $\theta_0 = \pm \pi/2$ much like spiral points trajectories in phase space are expelled outwards and also make open orbits in phase space, as shown in Fig \ref{fig:pstr0}-(a) and (b). Time signals for different initial conditions are qualitatively similar but are different showing sensitivity in initial conditions. In any case, observed orbits oscillate inside a region around the fixed-point. This region thins out as the Reynolds number decreases and converges into a stable limit cycle. Stability properties of these three fixed-points depend on the Reynolds number, $\theta_s = 0$ is a saddle point as long as there is vortex shedding behind the body. The other two fixed-points, $\theta_c = \pm \pi/2$, have an associated stable limit cycle at lower Reynolds number values. 

According to Williamson and Roshko's nomenclature~\cite{williamson88}, the wake produced in this regime is a 2S pattern. The vortices shed in each cycle seem to organize in a von K\'arm\'an wake close to the body, downstream vortex pairs interact with each other changing their pattern, as shown in figure \ref{fig:pstr0}-(c) where the elliptical cylinder at two points in time separated by one cycle is shown. Even though the two points in time seem to be similar, in each cycle the vorticity produced on the cylinder's surface changes slightly modifying the pattern downstream.  
     
\begin{figure}[ht]
\centering
  \includegraphics[width=0.8\linewidth]{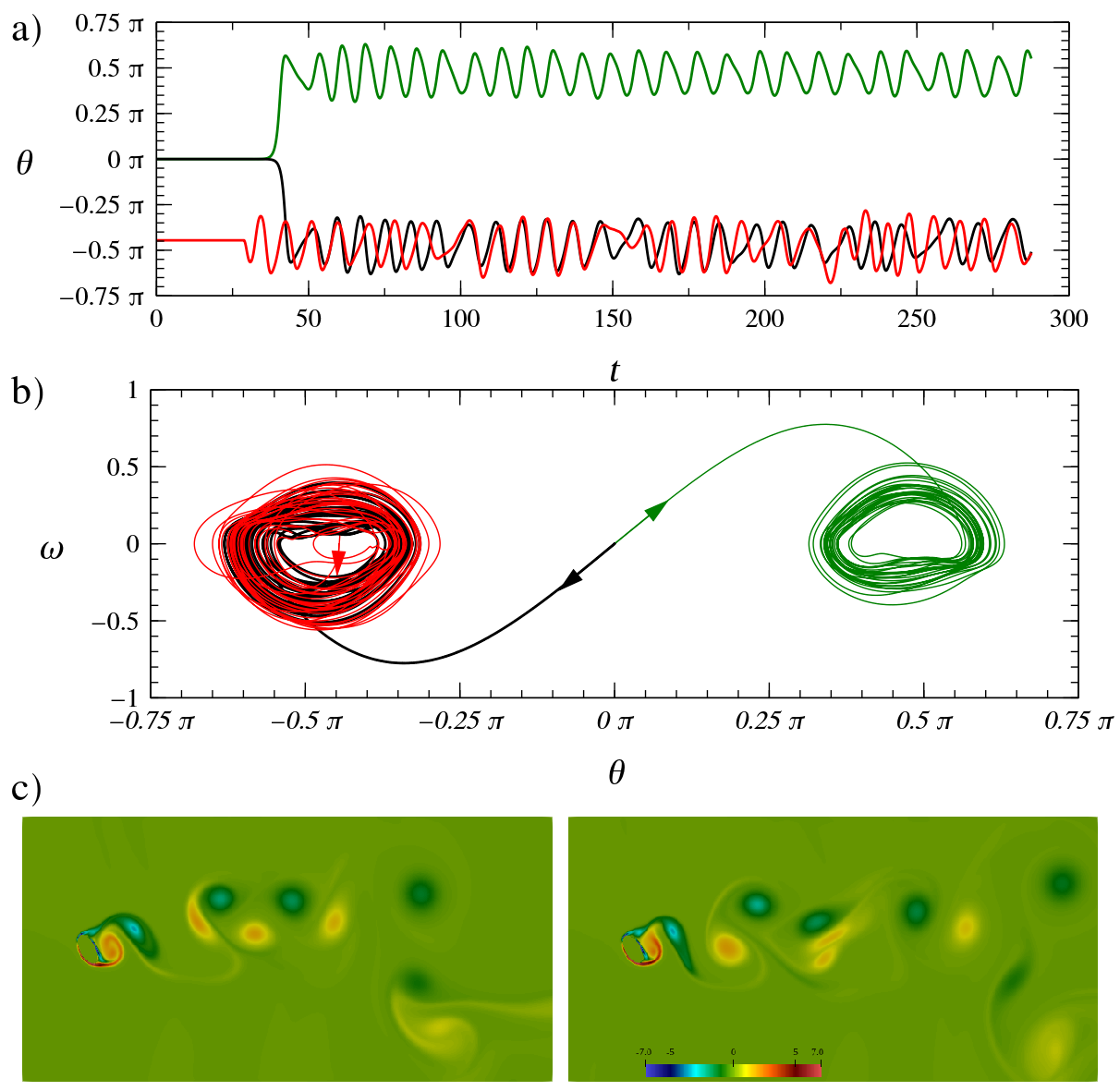}
  \caption{a) Rotational angle as function of time for different initial conditions with $S_t=0.4$. b) Phase space trajectories correspond to those shown in (a), arrows indicate the trajectories' starting point. c) Vorticity field at two different times separated approximately by one cycle.}
  \label{fig:pstr20}
\end{figure}%
Similar results were found by Weymouth~\cite{weymouth}, who studied a rotating ellipse towed with a constant velocity, when the axle of rotation passed through its geometrical center. In our case, the ellipse's response is a non periodic oscillation with an amplitude that varies in a complicated manner; showing that hydrodynamic torque is a non-linear function that plays the role of forcing and damping forces in the language of nonlinear oscillators~\cite{guckenheimer2013}. In VIV's literature, hydrodynamic forces due to vortex shedding processes are modeled as oscillatory periodic functions of time~\cite{sarpkaya,williamson}.

When torsional spring is included with its equilibrium position at $\theta_e=0$, there is a competition between restorative and hydrodynamic torques. When $0<S_t<0.8$, we found a similar behavior to the one described above (see figure \ref{fig:pstr20}-(a)). All initial conditions drive the elliptical cylinder to oscillate around one of two fixed points $ \pm \theta_c$ that have moved towards the point $\theta_s$ which is still a saddle point, as shown in figure \ref{fig:pstr20}-(b). As in the previous region, orbits are confined in regions that thin out as the Reynolds number decreases. Mean amplitude remains  approximately constant within this region, while the power spectral analysis of $\theta(t)$ shows several frequencies. The highest peak in power spectral density for each experiment (main frequency) is within the interval $[U/b,2U/b]$ and seems to change rather erratically due to the oscillation's non periodicity.

\begin{figure}[ht]
\centering
  \includegraphics[width=0.8\linewidth]{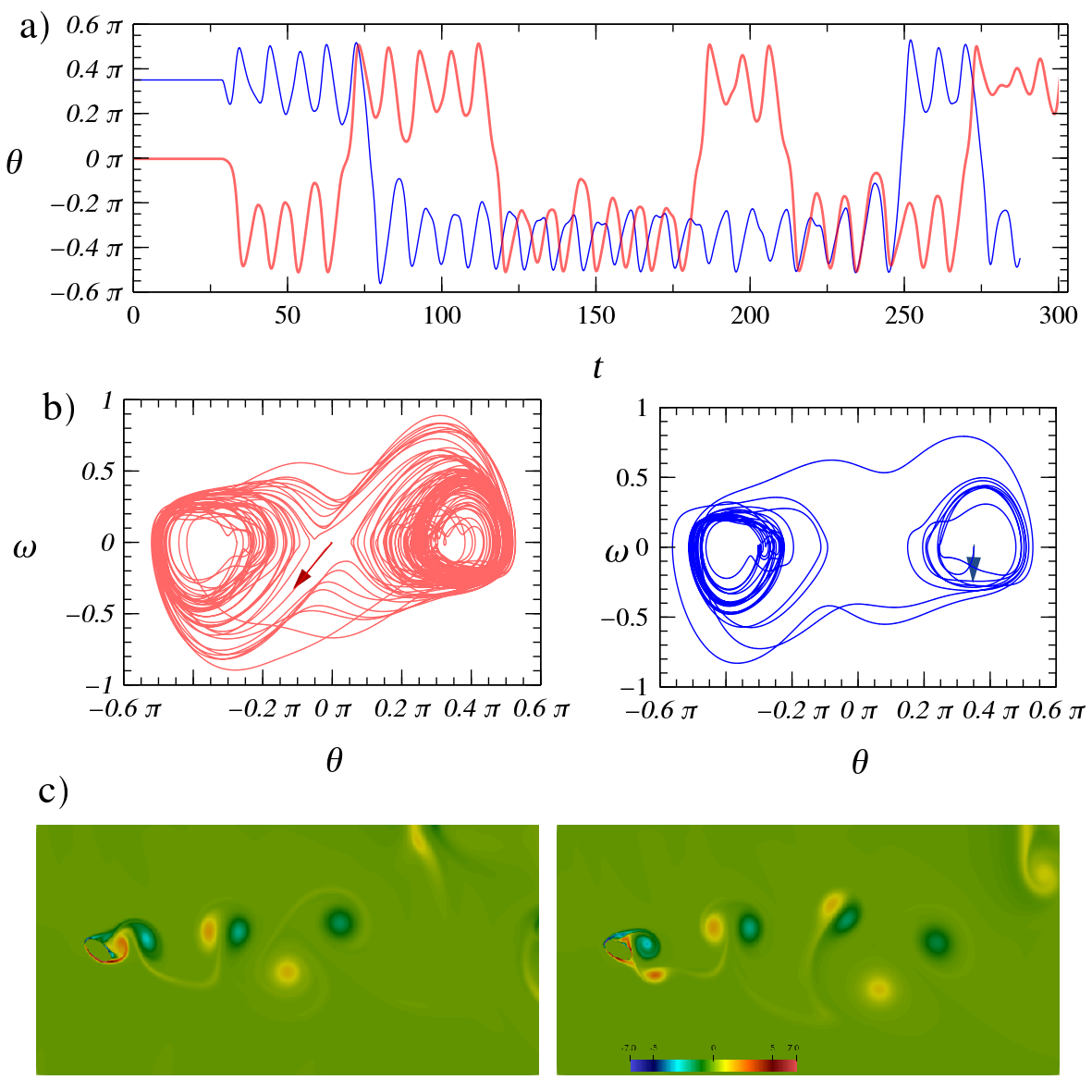}
  \caption{a) Rotational angle as function of time for different initial conditions with $S_t=0.84$. b) Phase space trajectories correspond to those shown in (a), the arrows indicate the trajectories' starting point. c) Vorticity field at two different times separated approximately by an oscillation of the elliptical cylinder.}
  \label{fig:pstr42}
\end{figure}%
In contrast with the case when $S_t=0$, symmetry properties of trajectories in phase space change. In the former case, phase space trajectories are symmetric with respect to the origin. When $S_t \neq 0$, trajectories  are skew-symmetric under a reflection in $\theta(t)$. The elliptical cylinder rotates faster when it moves away than when it moves towards the saddle point $\theta_s$. The vorticity field also shows different symmetrical properties, as shown in figure \ref{fig:pstr20}-(c), where the ellipse at approximately the same angular position after one cycle is shown. The vorticity generated at the cylinder's surface is rather different from one cycle to the next. Form, behavior and the way in which vortices interact with one another also change. They are deflected in the cross-flow direction which results in a wider wake.      
\begin{figure}[ht]
\centering
  \includegraphics[width=0.8\linewidth]{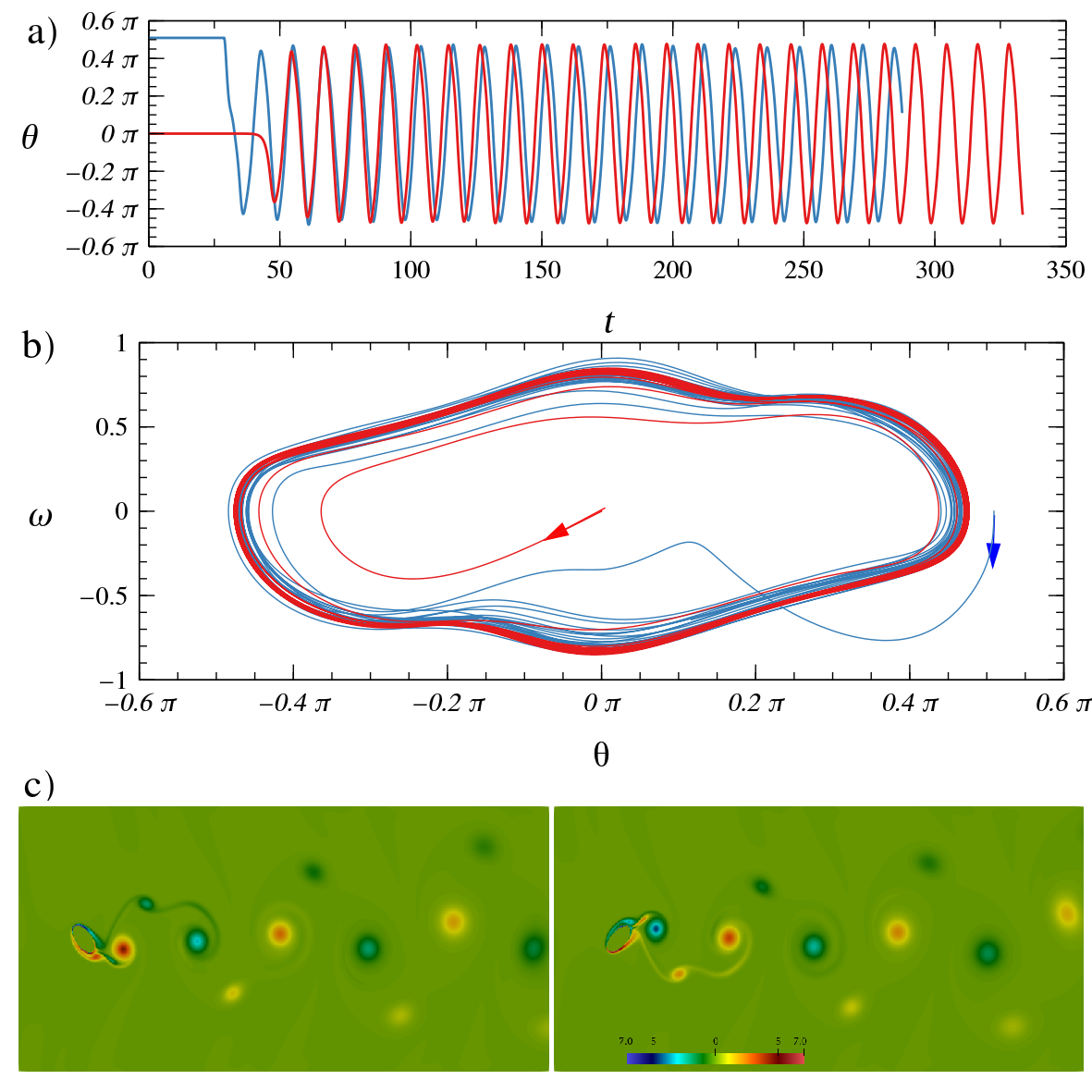}
  \caption{a) Rotational angle as function of time for different initial conditions for $S_t=1.16$. b) Phase space trajectories correspond to those shown in (a), arrows indicate the trajectories' starting point. c) Vorticity field at two different times separated approximately by half an oscillation of the elliptical cylinder.}
  \label{fig:pstr58}
\end{figure}%

When $0.5\leq S_t < 1.1$ the ellipse starts to oscillate around either $\theta_c$ or $-\theta_c$ depending on initial conditions as in the previous region. After a time interval, amplitude of oscillation increases resulting in the elliptical cylinder being attracted to the opposite fixed-point and starting to oscillate around it for another time interval, and so on (see figure \ref{fig:pstr42}-(a)-(c)). Residence times, where the ellipse oscillates around either of the fixed-points $\pm \theta_c$, seem to be distributed randomly. The two fixed-points $\pm \theta_c$ approach to $\theta_c=0$, as $S_t$ gets close to one, where they both merge with the saddle point to form what seems to be a strange attractor. Dynamics are very similar to the ones found with the Lorenz system and Duffing oscillator~\cite{guckenheimer2013}. In figure \ref{fig:pstr42}-(c) the vorticity field is shown at two points in time after one oscillation. In this case, it is found that shed vorticity differs and vortex pairs are further deflected vertically downstream compared to the previous region.   
\begin{figure}[ht]
\centering
  \includegraphics[width=0.8\linewidth]{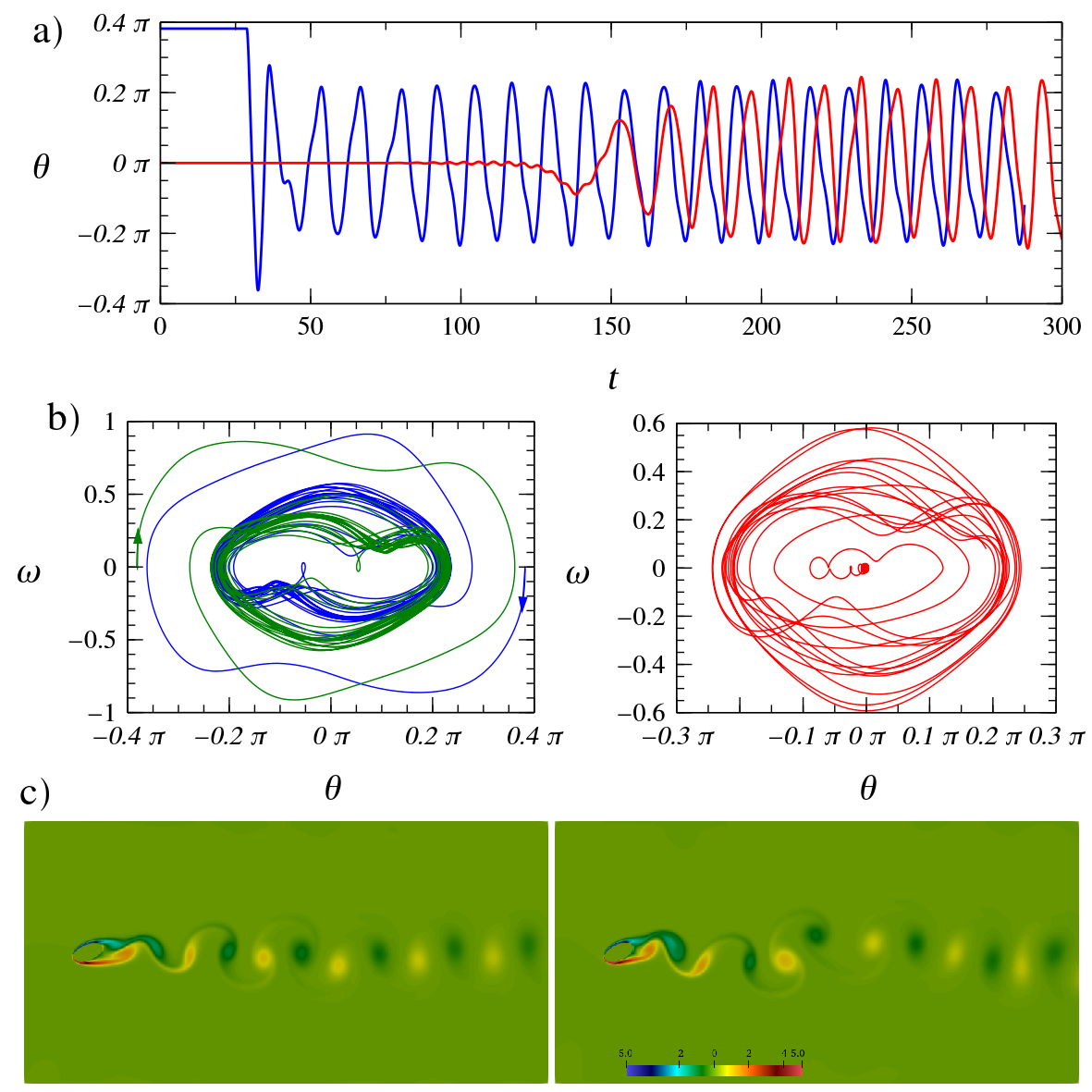}
  \caption{a) Rotational angle as function of time for different initial conditions for $S_t=1.68$. b) Phase space trajectories correspond to those shown in (a), arrows indicate trajectories' starting point. c) Vorticity field at two different times separated approximately by an oscillation of the elliptical cylinder.}
  \label{fig:pstr84}
\end{figure}%

For $1.1<S_t<1.2$ (shedding and spring frequencies are approximately the same) we found stable limit cycles, as shown in figure \ref{fig:pstr58}-(a) and (b). Observed $\theta(t)$ frequencies are approximately $U/2b$, half the shedding frequency, and amplitude of oscillation reaches its maximum values. Both lift and torque oscillate within the the same frequency as $\theta(t)$ and drag coefficient with shedding frequency $U/b$. Now, the only fixed-point seems to be $\theta_s = 0$ with a skew-symmetric stable limit cycle. Vorticity field is a 2P+2S wake, as shown in figure \ref{fig:pstr58}-(c) where we show two points in time separated by half a cycle of the ellipse's rotation. 

Notice that for each half a cycle, two structures are shed: a vortex dipole-like pair and a single one. The vortex pair is composed of two vortices with different strengths, the least intense one dissipates just before being shed and the most intense is dragged towards the center line. The single vortex shed travels to one side of the rest of the dipole-like vortex. The wake seems to be an internal inverted von K\'arm\'an street with an external one as shown in figure \ref{fig:pstr58}-(c). 

As $S_t$ is increased outside the previous region, orbits oscillate around a thicker region and the rotational angle loses its periodicity with amplitudes that vary from one cycle to the next by small amounts, as can be appreciated in figure \ref{fig:pstr84}-(a). In figure \ref{fig:pstr84}-(b) three types of trajectories are shown with different initial conditions. When the initial condition is $\pm \theta_0$ ($ \neq 0 $) we found two skew-symmetric solutions and, when the initial condition is close to $\theta_0 = 0$, solutions found seem to be a combination of the first two. Hence, even when it seems that fixed-points collapse at the saddle point position $\theta_s$, influence of the two fixed-points $\pm \theta_c$ can still be seen, which shows that basins of attraction are divided into at least three regions. 

Vorticity fields are made by 2S patterns, as in von K\'arm\'an wake, however, when angular velocity has changes as $\theta(t)$ approaches the saddle point (see figure \ref{fig:pstr84}-(b)) vortices are shed with slightly different amplitudes that modify the pattern downstream, as shown in figure \ref{fig:pstr84}-(c). In contrast with the initial region, where the amplitudes of oscillation around either of the fixed-points $\pm \theta_c $ remain approximately constant, in this region amplitude of oscillation rapidly decreases and, at some value of $S_t$, reaches a minimum value and a limit cycle is found. The cylinder's oscillations in this region are both very small and periodic with a frequency of two times the shedding frequency $U/b$.
\begin{figure}[ht]
\centering
  \includegraphics[width=1\linewidth]{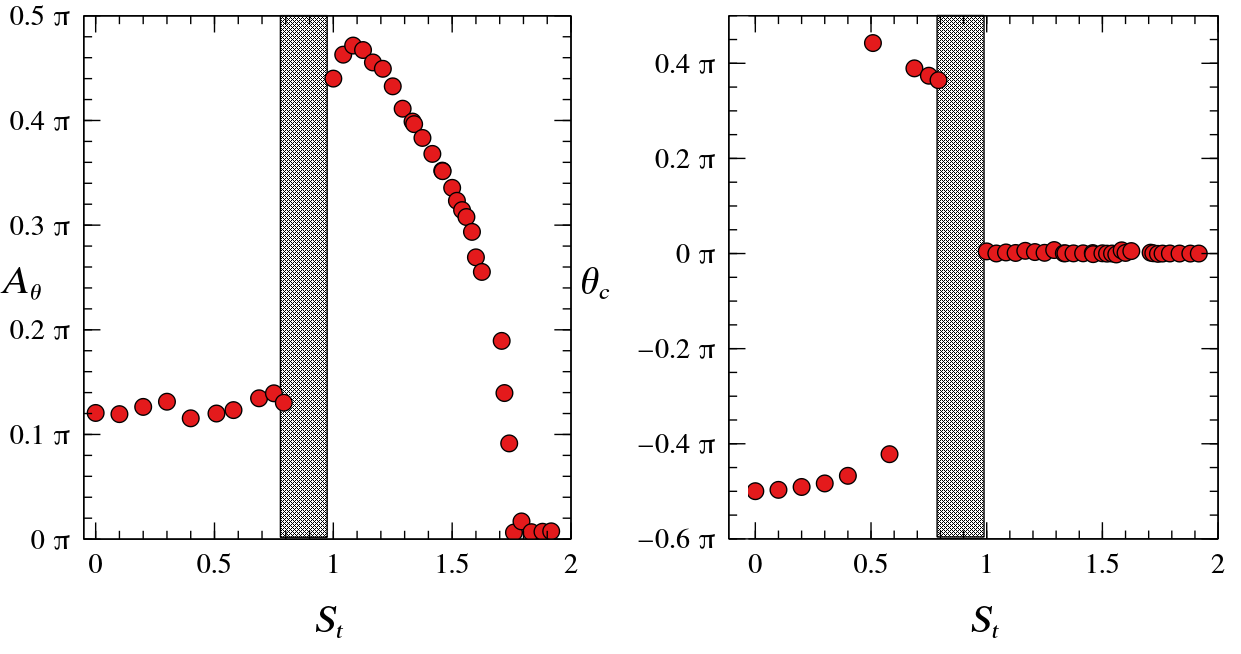}
  \caption{a) Root mean square of $\theta(t)$ as function of $S_t$. b) Evolution of $\pm \theta_c$ as function of $S_t$}
  \label{fig:rms}
\end{figure}%

Figure \ref{fig:rms} shows the root mean square of the angle of rotation as a function of $S_t$. As mentioned before, for small values of $S_t$, cylinder trajectories in phase space seem chaotic, variations in mean amplitudes of oscillation are small and remain approximately constant as $S_t$ grows. At some value of $S_t$, there is a transition where bigger oscillations are found, which correspond to time intervals when the cylinder passes from oscillating around one fixed-point to the other (shaded region in figure \ref{fig:rms}-(a)). Liapunov coefficients in these two regions are positive and have approximately the same value of $\lambda \sim 0.1$. The differences in initial conditions used to compute $\lambda$ were $10^{-7}$.

Evolution of fixed-points is shown in figure \ref{fig:rms}-(b). The starting point is $\theta_c = \pm \pi/2$ for $S_t=0$ and then $\theta_c$ decreases monotonically as $S_t$ grows. In the region of transition where the cylinder changes from chaotic oscillations around either of the fixed-points $\pm \theta_c$ to chaotic oscillations changing form one to the other intermittently, the behavior seems to follow the same trend. When $S_t \sim 1$ fixed-points jump and merge with the saddle point at $\theta_s=0$, marking the transition's boundary. 

The cylinder's amplitude of oscillation reaches its maximum when $1<S_t<1.2$, a tiny region where we found a stable limit cycle. The only fixed-point is at $\theta_s=0$, as shown in figure \ref{fig:rms}, the other two points collapse with $\theta_s$. However, their influence is still seen in the skew-symmetric phase space trajectories. As $S_t$ is increased, $A_{\theta}$ starts to decrease monotonically and phase space trajectories change their geometry, showing the influence of two fixed-points associated with hydrodynamic torque. When $S_t\sim 1.6$, amplitude $A_{\theta}$ decrease faster and becomes small but different from zero and remains constant. Trajectories in phase space are ellipses, oscillation is harmonic and, when $S_t \rightarrow \infty $, the only fixed-point becomes an attractor.

\section{Discussion}
\label{dis}
Non-linear oscillations of an elliptical cylinder that rotates around its axle due to flow of a Newtonian viscous fluid and a restorative force are discussed. When the only torque on the elliptical cylinder is due to fluid flow, the ellipse's major-axis oscillates around the axis in the cross-flow direction. Without fluid flow, the ellipse rotates harmonically around $\theta=0$ with a frequency of $\sqrt{k/I}$. Reynolds number used was $Re=200$ where it is known that hydrodynamic torque produces non-linear oscillations. We used $S_t$ as a central parameter, which results from a comparison between the characteristic vortex shedding $U/b$ and spring $\sqrt{k/I}$ frequencies.
\begin{figure}[ht]
\centering
  \includegraphics[width=1\linewidth]{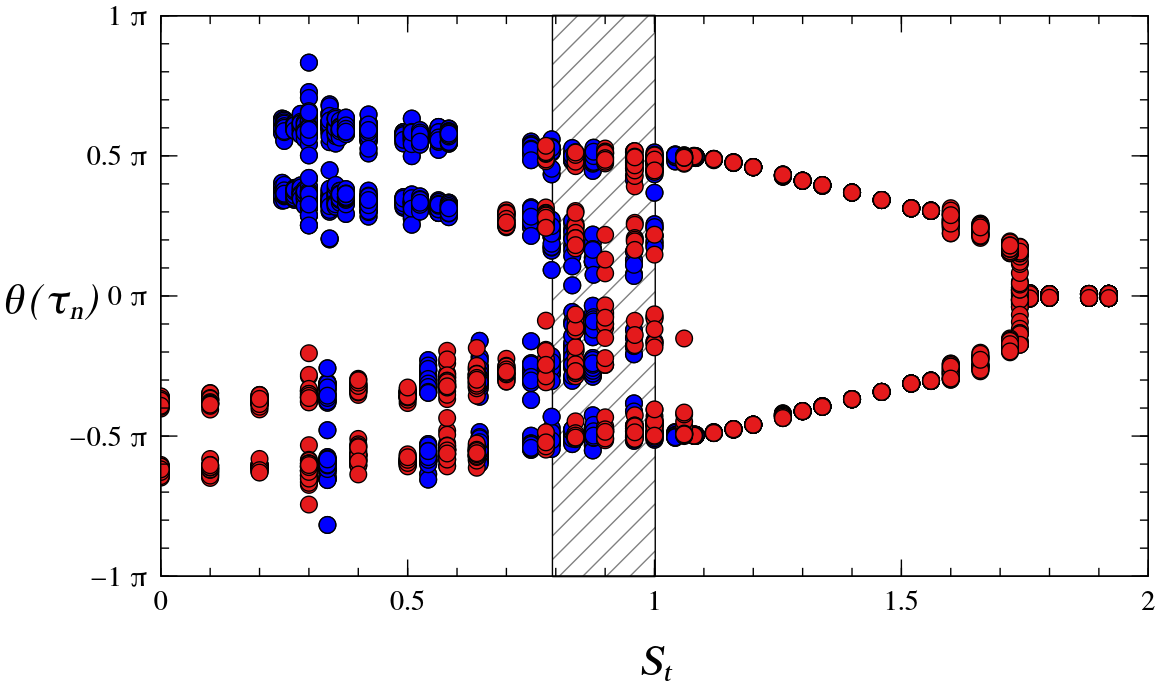}
  \caption{Bifurcation diagram. The values of $\tau_n$ are such that $\omega(\tau_n)=0$. Colors represent different initial conditions.}
  \label{fig:bif}
\end{figure}%

Results found showed similar behavior to vortex induced vibrations (VIV) around a circular cylinder with two and three degrees of freedom~\cite{govardhan2000,brika,williamson,reyes}. Behavior of oscillation amplitudes in the cross-flow direction is classified into three regions as a function of reduced velocity, which is proportional to the inverse of $S_t$. The excitation region is where the cross-flow amplitude of oscillation is small but starts to increase. The lock-in and desynchronization regions are regions where the cross-flow amplitude reaches its maximum and begins to decrease, respectively.  

In our case, the excitation branch can be identified as the region when $S_t>1.8$, as shown in figure \ref{fig:bif}, where a bifurcation diagram for this problem is portrayed. The fixed-point $\theta_s=0$ behaves as an attractor when $S_t\rightarrow \infty$ (not shown on bifurcation diagram). Hydrodynamic torque works as a damping term in equation (\ref{nec}) since every initial condition leads the ellipse to a rest, as in a damped pendulum. As $S_t$ decreases, the ellipse starts to oscillate periodically with a period of $2b/U$ and an amplitude that remains approximately constant, similar to a forced damped pendulum. 

The excitation branch is followed by an abrupt growing of amplitudes of oscillation which lead to the lock-in branch, where the cross-flow amplitude reaches its maximum. In our case, as $S_t$ decreases, amplitude starts to grow abruptly and reaches a maximum where we found a stable limit cycle. Around the maxima, we found a region where oscillation is periodic, which we identified as the lock-in region. Transition between excitation and lock-in regions occurs in a region where the stable limit cycle evolves in a new structure influenced by two fixed-points at $\pm \theta_c$. Amplitude reached at each oscillation varies, producing asymmetries in the branches of the parabola-like curve in figure \ref{fig:bif}. As $S_t$ decreases, observed variations in amplitude diminishes and reaches zero. As $S_t$ approaches one, amplitudes collapse to a single value with a more uniform growth than in the transition region.    

After the lock-in branch, the stable limit cycle evolves in what seems to be a strange attractor (shaded region in figure \ref{fig:bif}). Even when it seems that amplitudes of $\theta(t)$ still explore a wide range of values, amplitude of oscillation decreased since the ellipse oscillates intermittently about either $\theta_c$ or $-\theta_c$. After the transition region, amplitudes of oscillation remain approximately constant and fixed-points $\pm \theta_c$ increase towards $\pm \pi/2$ as $S_t$ approaches zero. We identified this region as the desynchronization zone. However, in our case, the ellipse's oscillation is non-periodic and its amplitude is larger than in the excitation branch. 

Within the transition region the ellipse oscillates intermittently around fixed-points $\pm \theta_c$. In the language of VIV's on circular cylinders, this region of transition is identified with a hysteresis process, meaning that it is different to go from the excitation region to the lock-in region than the other way around~\cite{brika,williamson,Singh}. There is also evidence in 2D numerical simulations that transition between these two regions can be made following different paths by changing the set of initial conditions, because there are several coexisting solutions between the excitation region and the lock-in branch~\cite{reyes}.

The system under consideration shows a close behavior with the problem of VIV's with 2 degrees of freedom in a sense that similar regions or branches can be identified. However, in this case, we only found periodic solutions in a small region (identified as the lock-in branch) and when $S_t \rightarrow \infty$ which corresponds to a fixed ellipse. When $S_t=0$ the ellipse's oscillation is chaotic due to the non-linear nature of hydrodynamic torque. As $S_t$ is increased, degree of confinement in cylinder movement is modified and shed vorticity changes, which is due to a competition between hydrodynamic and restorative torques. Therefore, only when both torques synchronize or the movement is confined to very small oscillations, obtained solutions are periodic.  In the rest of analyzed regions, the hydrodynamic torque dominates the flow and the ellipse's oscillation shows chaotic dynamics. For a future works, we will be extending the study to ellipses with three degrees of freedom and will include other bodies, to study solid-fluid-solid hydrodynamic interactions.

\bibliography{bibl-lbe}

\begin{thebibliography}{28}%
\makeatletter
\providecommand \@ifxundefined [1]{%
 \@ifx{#1\undefined}
}%
\providecommand \@ifnum [1]{%
 \ifnum #1\expandafter \@firstoftwo
 \else \expandafter \@secondoftwo
 \fi
}%
\providecommand \@ifx [1]{%
 \ifx #1\expandafter \@firstoftwo
 \else \expandafter \@secondoftwo
 \fi
}%
\providecommand \natexlab [1]{#1}%
\providecommand \enquote  [1]{``#1''}%
\providecommand \bibnamefont  [1]{#1}%
\providecommand \bibfnamefont [1]{#1}%
\providecommand \citenamefont [1]{#1}%
\providecommand \href@noop [0]{\@secondoftwo}%
\providecommand \href [0]{\begingroup \@sanitize@url \@href}%
\providecommand \@href[1]{\@@startlink{#1}\@@href}%
\providecommand \@@href[1]{\endgroup#1\@@endlink}%
\providecommand \@sanitize@url [0]{\catcode `\\12\catcode `\$12\catcode
  `\&12\catcode `\#12\catcode `\^12\catcode `\_12\catcode `\%12\relax}%
\providecommand \@@startlink[1]{}%
\providecommand \@@endlink[0]{}%
\providecommand \url  [0]{\begingroup\@sanitize@url \@url }%
\providecommand \@url [1]{\endgroup\@href {#1}{\urlprefix }}%
\providecommand \urlprefix  [0]{URL }%
\providecommand \Eprint [0]{\href }%
\providecommand \doibase [0]{http://dx.doi.org/}%
\providecommand \selectlanguage [0]{\@gobble}%
\providecommand \bibinfo  [0]{\@secondoftwo}%
\providecommand \bibfield  [0]{\@secondoftwo}%
\providecommand \translation [1]{[#1]}%
\providecommand \BibitemOpen [0]{}%
\providecommand \bibitemStop [0]{}%
\providecommand \bibitemNoStop [0]{.\EOS\space}%
\providecommand \EOS [0]{\spacefactor3000\relax}%
\providecommand \BibitemShut  [1]{\csname bibitem#1\endcsname}%
\let\auto@bib@innerbib\@empty
\bibitem [{\citenamefont {Williamson}\ and\ \citenamefont
  {Govardhan}(2004)}]{williamson}%
  \BibitemOpen
  \bibfield  {author} {\bibinfo {author} {\bibfnamefont {C.~H.~K.}\
  \bibnamefont {Williamson}}\ and\ \bibinfo {author} {\bibfnamefont
  {R.}~\bibnamefont {Govardhan}},\ }\bibfield  {title} {\enquote {\bibinfo
  {title} {{Vortex-induced vibrations}},}\ }\href@noop {} {\bibfield  {journal}
  {\bibinfo  {journal} {Annu. Rev. Fluid Mech.}\ }\textbf {\bibinfo {volume}
  {36}},\ \bibinfo {pages} {413--455} (\bibinfo {year} {2004})}\BibitemShut
  {NoStop}%
\bibitem [{\citenamefont {Sarpkaya}(2004)}]{sarpkaya}%
  \BibitemOpen
  \bibfield  {author} {\bibinfo {author} {\bibfnamefont {T.}~\bibnamefont
  {Sarpkaya}},\ }\bibfield  {title} {\enquote {\bibinfo {title} {{A Critical
  review of the intrinsic nature of vortex-induced vibrations}},}\ }\href@noop
  {} {\bibfield  {journal} {\bibinfo  {journal} {J. Fluids Struct.}\ }\textbf
  {\bibinfo {volume} {19}},\ \bibinfo {pages} {389--447} (\bibinfo {year}
  {2004})}\BibitemShut {NoStop}%
\bibitem [{\citenamefont {Bearman}(1984)}]{bearman1984}%
  \BibitemOpen
  \bibfield  {author} {\bibinfo {author} {\bibfnamefont {P.~W.}\ \bibnamefont
  {Bearman}},\ }\bibfield  {title} {\enquote {\bibinfo {title} {{Vortex
  shedding from oscillating bluff bodies}},}\ }\href@noop {} {\bibfield
  {journal} {\bibinfo  {journal} {Rev. Fluid Mech.}\ }\textbf {\bibinfo
  {volume} {16}},\ \bibinfo {pages} {195--222} (\bibinfo {year}
  {1984})}\BibitemShut {NoStop}%
\bibitem [{\citenamefont {Bhattacharya}\ and\ \citenamefont
  {Sorathiya-Shahajhan}(2016)}]{bhattacharya16}%
  \BibitemOpen
  \bibfield  {author} {\bibinfo {author} {\bibfnamefont {A.}~\bibnamefont
  {Bhattacharya}}\ and\ \bibinfo {author} {\bibfnamefont {S.~S.}\ \bibnamefont
  {Sorathiya-Shahajhan}},\ }\bibfield  {title} {\enquote {\bibinfo {title}
  {Power extration form vortex-induced angular oscillations of elliptical
  cylinder},}\ }\href@noop {} {\bibfield  {journal} {\bibinfo  {journal}
  {Journal of Fluids and Structures.}\ }\textbf {\bibinfo {volume} {63}},\
  \bibinfo {pages} {140--154} (\bibinfo {year} {2016})}\BibitemShut {NoStop}%
\bibitem [{\citenamefont {Abu~Shahzer}\ \emph {et~al.}(2022)\citenamefont
  {Abu~Shahzer}, \citenamefont {Athar~Khan}, \citenamefont {Fahad~Anwer},
  \citenamefont {Anwar~Khan}, \citenamefont {Shoaib~Khan}, \citenamefont
  {Algethami},\ and\ \citenamefont {M.}}]{abu22}%
  \BibitemOpen
  \bibfield  {author} {\bibinfo {author} {\bibfnamefont {M.}~\bibnamefont
  {Abu~Shahzer}}, \bibinfo {author} {\bibfnamefont {M.}~\bibnamefont
  {Athar~Khan}}, \bibinfo {author} {\bibfnamefont {S.}~\bibnamefont
  {Fahad~Anwer}}, \bibinfo {author} {\bibfnamefont {S.}~\bibnamefont
  {Anwar~Khan}}, \bibinfo {author} {\bibfnamefont {M.}~\bibnamefont
  {Shoaib~Khan}}, \bibinfo {author} {\bibfnamefont {A.}~\bibnamefont
  {Algethami}}, \ and\ \bibinfo {author} {\bibfnamefont {A.}~\bibnamefont
  {M.}},\ }\bibfield  {title} {\enquote {\bibinfo {title} {A comprehensive
  investigation of vortex-induced vibrations and flow-induced rotation of an
  elliptic cylinder},}\ }\href {\doibase doi: 10.1063/5.0079642} {\bibfield
  {journal} {\bibinfo  {journal} {Phys. Fluids}\ }\textbf {\bibinfo {volume}
  {34}},\ \bibinfo {pages} {033605} (\bibinfo {year} {2022})}\BibitemShut
  {NoStop}%
\bibitem [{\citenamefont {Navrose}\ \emph {et~al.}(2014)\citenamefont
  {Navrose}, \citenamefont {Yogeswaran}, \citenamefont {Sen},\ and\
  \citenamefont {Mittal}}]{NAVROSE201455}%
  \BibitemOpen
  \bibfield  {author} {\bibinfo {author} {\bibnamefont {Navrose}}, \bibinfo
  {author} {\bibfnamefont {V.}~\bibnamefont {Yogeswaran}}, \bibinfo {author}
  {\bibfnamefont {S.}~\bibnamefont {Sen}}, \ and\ \bibinfo {author}
  {\bibfnamefont {S.}~\bibnamefont {Mittal}},\ }\bibfield  {title} {\enquote
  {\bibinfo {title} {Free vibrations of an elliptic cylinder at low reynolds
  numbers},}\ }\href {\doibase
  https://doi.org/10.1016/j.jfluidstructs.2014.07.012} {\bibfield  {journal}
  {\bibinfo  {journal} {Journal of Fluids and Structures}\ }\textbf {\bibinfo
  {volume} {51}},\ \bibinfo {pages} {55--67} (\bibinfo {year}
  {2014})}\BibitemShut {NoStop}%
\bibitem [{\citenamefont {Newman}\ and\ \citenamefont
  {Karnidakis}(1997)}]{newman97}%
  \BibitemOpen
  \bibfield  {author} {\bibinfo {author} {\bibfnamefont {D.~J.}\ \bibnamefont
  {Newman}}\ and\ \bibinfo {author} {\bibfnamefont {G.~E.}\ \bibnamefont
  {Karnidakis}},\ }\bibfield  {title} {\enquote {\bibinfo {title} {A direct
  numerical simulation study of flow past a freely vibrating cable},}\
  }\href@noop {} {\bibfield  {journal} {\bibinfo  {journal} {J. Fluid Mech.}\
  }\textbf {\bibinfo {volume} {344}},\ \bibinfo {pages} {95--136} (\bibinfo
  {year} {1997})}\BibitemShut {NoStop}%
\bibitem [{\citenamefont {Brika}\ and\ \citenamefont
  {Laneville}(1993)}]{brika}%
  \BibitemOpen
  \bibfield  {author} {\bibinfo {author} {\bibfnamefont {D.}~\bibnamefont
  {Brika}}\ and\ \bibinfo {author} {\bibfnamefont {A.}~\bibnamefont
  {Laneville}},\ }\bibfield  {title} {\enquote {\bibinfo {title}
  {{Vortex-induced vibrations of a long flexible circular cylinder}},}\
  }\href@noop {} {\bibfield  {journal} {\bibinfo  {journal} {J. Fluid Mech.}\
  }\textbf {\bibinfo {volume} {250}},\ \bibinfo {pages} {481--508} (\bibinfo
  {year} {1993})}\BibitemShut {NoStop}%
\bibitem [{\citenamefont {Yamamoto}\ \emph {et~al.}(2004)\citenamefont
  {Yamamoto}, \citenamefont {Meneghini}, \citenamefont {Saltara}, \citenamefont
  {Fregonesi},\ and\ \citenamefont {Ferrari}}]{yamamoto04}%
  \BibitemOpen
  \bibfield  {author} {\bibinfo {author} {\bibfnamefont {C.~T.}\ \bibnamefont
  {Yamamoto}}, \bibinfo {author} {\bibfnamefont {J.~R.}\ \bibnamefont
  {Meneghini}}, \bibinfo {author} {\bibfnamefont {F.}~\bibnamefont {Saltara}},
  \bibinfo {author} {\bibfnamefont {R.~A.}\ \bibnamefont {Fregonesi}}, \ and\
  \bibinfo {author} {\bibfnamefont {J.~A.}\ \bibnamefont {Ferrari}},\
  }\bibfield  {title} {\enquote {\bibinfo {title} {Numerical simulations of
  vortex-induced vibration on flexible cylinders},}\ }\href@noop {} {\bibfield
  {journal} {\bibinfo  {journal} {Journal of Fluids and Structures.}\ }\textbf
  {\bibinfo {volume} {19}},\ \bibinfo {pages} {467--489} (\bibinfo {year}
  {2004})}\BibitemShut {NoStop}%
\bibitem [{\citenamefont {Jauvtis}\ and\ \citenamefont
  {Williamson}(2003)}]{jauvtis03}%
  \BibitemOpen
  \bibfield  {author} {\bibinfo {author} {\bibfnamefont {N.}~\bibnamefont
  {Jauvtis}}\ and\ \bibinfo {author} {\bibfnamefont {C.~H.~K.}\ \bibnamefont
  {Williamson}},\ }\bibfield  {title} {\enquote {\bibinfo {title}
  {{Vortex-induced vibration of a cylinder with two degrees of freedom}},}\
  }\href@noop {} {\bibfield  {journal} {\bibinfo  {journal} {J. Fluids
  Struct.}\ }\textbf {\bibinfo {volume} {17}},\ \bibinfo {pages} {1035--1042}
  (\bibinfo {year} {2003})}\BibitemShut {NoStop}%
\bibitem [{\citenamefont {Reyes}\ and\ \citenamefont
  {Mandujano}(2022)}]{reyes}%
  \BibitemOpen
  \bibfield  {author} {\bibinfo {author} {\bibfnamefont {M.}~\bibnamefont
  {Reyes}}\ and\ \bibinfo {author} {\bibfnamefont {F.}~\bibnamefont
  {Mandujano}},\ }\bibfield  {title} {\enquote {\bibinfo {title} {{Vortex
  induced vibrations of a cylinder at low mass ratio}},}\ }\href@noop {}
  {\bibfield  {journal} {\bibinfo  {journal} {Eur. J. Mech. /B Fluids}\
  }\textbf {\bibinfo {volume} {91}},\ \bibinfo {pages} {66--79} (\bibinfo
  {year} {2022})}\BibitemShut {NoStop}%
\bibitem [{\citenamefont {Dahl}, \citenamefont {Hover},\ and\ \citenamefont
  {Triantafyllou}(2006)}]{dahl06}%
  \BibitemOpen
  \bibfield  {author} {\bibinfo {author} {\bibfnamefont {J.~M.}\ \bibnamefont
  {Dahl}}, \bibinfo {author} {\bibfnamefont {M.~S.}\ \bibnamefont {Hover}}, \
  and\ \bibinfo {author} {\bibfnamefont {M.~S.}\ \bibnamefont
  {Triantafyllou}},\ }\bibfield  {title} {\enquote {\bibinfo {title}
  {Two-degree-of-freedom vortex-induced vibrations using a force assisted
  apparatus},}\ }\href@noop {} {\bibfield  {journal} {\bibinfo  {journal}
  {Journal of Fluids and Structures.}\ }\textbf {\bibinfo {volume} {22}},\
  \bibinfo {pages} {807--818} (\bibinfo {year} {2006})}\BibitemShut {NoStop}%
\bibitem [{\citenamefont {Weymouth}\ and\ \citenamefont
  {Yue}(2011)}]{weymouth}%
  \BibitemOpen
  \bibfield  {author} {\bibinfo {author} {\bibfnamefont {G.~D.}\ \bibnamefont
  {Weymouth}}\ and\ \bibinfo {author} {\bibfnamefont {D.~K.~P.}\ \bibnamefont
  {Yue}},\ }\bibfield  {title} {\enquote {\bibinfo {title} {{Boundary data
  immersion method for cartesian-grid simulations of fluid-body interaction
  problems}},}\ }\href@noop {} {\bibfield  {journal} {\bibinfo  {journal} {J.
  Comput. Phys.}\ }\textbf {\bibinfo {volume} {230}},\ \bibinfo {pages}
  {6233--6247} (\bibinfo {year} {2011})}\BibitemShut {NoStop}%
\bibitem [{\citenamefont {Ros\'en}(2017)}]{rosen2017}%
  \BibitemOpen
  \bibfield  {author} {\bibinfo {author} {\bibfnamefont {T.}~\bibnamefont
  {Ros\'en}},\ }\bibfield  {title} {\enquote {\bibinfo {title} {Chaotic
  rotation of a spheroidal particle in simple shear flow},}\ }\href@noop {}
  {\bibfield  {journal} {\bibinfo  {journal} {Chaos}\ }\textbf {\bibinfo
  {volume} {27}},\ \bibinfo {pages} {063112} (\bibinfo {year}
  {2017})}\BibitemShut {NoStop}%
\bibitem [{\citenamefont {Zhao}, \citenamefont {Leontini},\ and\ \citenamefont
  {J.}(2014)}]{zhao14}%
  \BibitemOpen
  \bibfield  {author} {\bibinfo {author} {\bibfnamefont {J.}~\bibnamefont
  {Zhao}}, \bibinfo {author} {\bibfnamefont {D.~L.~J.}\ \bibnamefont
  {Leontini}}, \ and\ \bibinfo {author} {\bibfnamefont {S.}~\bibnamefont
  {J.}},\ }\bibfield  {title} {\enquote {\bibinfo {title} {Chaotic vortex
  induced vibrations},}\ }\href@noop {} {\bibfield  {journal} {\bibinfo
  {journal} {Physics of Fluids}\ }\textbf {\bibinfo {volume} {26}},\ \bibinfo
  {pages} {121702} (\bibinfo {year} {2014})}\BibitemShut {NoStop}%
\bibitem [{\citenamefont {Maxwell}(2011)}]{maxwell}%
  \BibitemOpen
  \bibfield  {author} {\bibinfo {author} {\bibfnamefont {J.~C.}\ \bibnamefont
  {Maxwell}},\ }\enquote {\bibinfo {title} {On a particular case of the descent
  of a heavy body in a resisting medium},}\ in\ \href {\doibase
  10.1017/CBO9780511698095.008} {\emph {\bibinfo {booktitle} {The Scientific
  Papers of James Clerk Maxwell}}},\ \bibinfo {series} {Cambridge Library
  Collection - Physical Sciences}, Vol.~\bibinfo {volume} {1}\ (\bibinfo
  {publisher} {Cambridge University Press},\ \bibinfo {year} {2011})\ p.\
  \bibinfo {pages} {115–118}\BibitemShut {NoStop}%
\bibitem [{\citenamefont {Lugt}(1980)}]{lugt}%
  \BibitemOpen
  \bibfield  {author} {\bibinfo {author} {\bibfnamefont {H.~J.}\ \bibnamefont
  {Lugt}},\ }\bibfield  {title} {\enquote {\bibinfo {title} {{Autorotation of
  an elliptic cylinder about an axis perpendicular to the flow}},}\ }\href@noop
  {} {\bibfield  {journal} {\bibinfo  {journal} {J. Fluid Mech.}\ }\textbf
  {\bibinfo {volume} {94-4}},\ \bibinfo {pages} {817--840} (\bibinfo {year}
  {1980})}\BibitemShut {NoStop}%
\bibitem [{\citenamefont {Lugt}(1983)}]{lugt83}%
  \BibitemOpen
  \bibfield  {author} {\bibinfo {author} {\bibfnamefont {H.~J.}\ \bibnamefont
  {Lugt}},\ }\bibfield  {title} {\enquote {\bibinfo {title} {Autorotation},}\
  }\href {\doibase 10.1146/annurev.fl.15.010183.001011} {\bibfield  {journal}
  {\bibinfo  {journal} {Annual Review of Fluid Mechanics}\ }\textbf {\bibinfo
  {volume} {15}},\ \bibinfo {pages} {123--147} (\bibinfo {year} {1983})},\
  \Eprint
  {http://arxiv.org/abs/https://doi.org/10.1146/annurev.fl.15.010183.001011}
  {https://doi.org/10.1146/annurev.fl.15.010183.001011} \BibitemShut {NoStop}%
\bibitem [{\citenamefont {He}, \citenamefont {Chen},\ and\ \citenamefont
  {Doolen}(1998)}]{He98}%
  \BibitemOpen
  \bibfield  {author} {\bibinfo {author} {\bibfnamefont {X.}~\bibnamefont
  {He}}, \bibinfo {author} {\bibfnamefont {S.}~\bibnamefont {Chen}}, \ and\
  \bibinfo {author} {\bibfnamefont {G.~D.}\ \bibnamefont {Doolen}},\ }\bibfield
   {title} {\enquote {\bibinfo {title} {{A Novel Thermal Model for the Lattice
  {B}oltzmann Method in Incompressible Limit}},}\ }\href@noop {} {\bibfield
  {journal} {\bibinfo  {journal} {Journal of Computational Physics}\ }\textbf
  {\bibinfo {volume} {146}},\ \bibinfo {pages} {282--300} (\bibinfo {year}
  {1998})}\BibitemShut {NoStop}%
\bibitem [{\citenamefont {Mandujano}\ and\ \citenamefont
  {M{\'a}laga}(2018)}]{mandujano18}%
  \BibitemOpen
  \bibfield  {author} {\bibinfo {author} {\bibfnamefont {M.}~\bibnamefont
  {Mandujano}}\ and\ \bibinfo {author} {\bibfnamefont {C.}~\bibnamefont
  {M{\'a}laga}},\ }\bibfield  {title} {\enquote {\bibinfo {title} {{On the
  forced flow around a rigid flapping foil}},}\ }\href@noop {} {\bibfield
  {journal} {\bibinfo  {journal} {Phys. Fluids}\ }\textbf {\bibinfo {volume}
  {30}},\ \bibinfo {pages} {061901} (\bibinfo {year} {2018})}\BibitemShut
  {NoStop}%
\bibitem [{\citenamefont {Guo}\ and\ \citenamefont {Zheng}(2002)}]{guo02a}%
  \BibitemOpen
  \bibfield  {author} {\bibinfo {author} {\bibfnamefont {Z.}~\bibnamefont
  {Guo}}\ and\ \bibinfo {author} {\bibfnamefont {C.}~\bibnamefont {Zheng}},\
  }\bibfield  {title} {\enquote {\bibinfo {title} {{{A}n extrapolation method
  for boundary conditions in lattice {B}oltzmann method}},}\ }\href@noop {}
  {\bibfield  {journal} {\bibinfo  {journal} {Phys. Fluids}\ }\textbf {\bibinfo
  {volume} {14}},\ \bibinfo {pages} {2007--2010} (\bibinfo {year}
  {2002})}\BibitemShut {NoStop}%
\bibitem [{\citenamefont {Mei}\ \emph {et~al.}(2002)\citenamefont {Mei},
  \citenamefont {Yu}, \citenamefont {Shyy},\ and\ \citenamefont {Luo}}]{mei02}%
  \BibitemOpen
  \bibfield  {author} {\bibinfo {author} {\bibfnamefont {R.}~\bibnamefont
  {Mei}}, \bibinfo {author} {\bibfnamefont {D.}~\bibnamefont {Yu}}, \bibinfo
  {author} {\bibfnamefont {W.}~\bibnamefont {Shyy}}, \ and\ \bibinfo {author}
  {\bibfnamefont {L.}~\bibnamefont {Luo}},\ }\bibfield  {title} {\enquote
  {\bibinfo {title} {{{F}orce evaluation in the lattice {B}oltzmann method
  involving curved geometry}},}\ }\href@noop {} {\bibfield  {journal} {\bibinfo
   {journal} {Phys. Rev. E}\ }\textbf {\bibinfo {volume} {65}},\ \bibinfo
  {pages} {041203} (\bibinfo {year} {2002})}\BibitemShut {NoStop}%
\bibitem [{\citenamefont {Mandujano}\ and\ \citenamefont
  {Rechtman}(2008)}]{mandujano08}%
  \BibitemOpen
  \bibfield  {author} {\bibinfo {author} {\bibfnamefont {M.}~\bibnamefont
  {Mandujano}}\ and\ \bibinfo {author} {\bibfnamefont {R.}~\bibnamefont
  {Rechtman}},\ }\bibfield  {title} {\enquote {\bibinfo {title} {{Thermal
  levitation}},}\ }\href@noop {} {\bibfield  {journal} {\bibinfo  {journal} {J.
  Fluid Mech.}\ }\textbf {\bibinfo {volume} {606}},\ \bibinfo {pages}
  {105--114} (\bibinfo {year} {2008})}\BibitemShut {NoStop}%
\bibitem [{\citenamefont {Weymouth}(2014)}]{weymouth2014}%
  \BibitemOpen
  \bibfield  {author} {\bibinfo {author} {\bibfnamefont {G.~D.}\ \bibnamefont
  {Weymouth}},\ }\bibfield  {title} {\enquote {\bibinfo {title} {{Chaotic
  rotation of a towed elliptical cylinder}},}\ }\href {\doibase
  10.1017/jfm.2014.42} {\bibfield  {journal} {\bibinfo  {journal} {J. Fluid
  Mech.}\ }\textbf {\bibinfo {volume} {743}},\ \bibinfo {pages} {385--398}
  (\bibinfo {year} {2014})}\BibitemShut {NoStop}%
\bibitem [{\citenamefont {Williamson}\ and\ \citenamefont
  {Roshko}(1988)}]{williamson88}%
  \BibitemOpen
  \bibfield  {author} {\bibinfo {author} {\bibfnamefont {C.~H.~K.}\
  \bibnamefont {Williamson}}\ and\ \bibinfo {author} {\bibfnamefont
  {A.}~\bibnamefont {Roshko}},\ }\bibfield  {title} {\enquote {\bibinfo {title}
  {{Vortex formation in the wake of an oscillating cylinder}},}\ }\href@noop {}
  {\bibfield  {journal} {\bibinfo  {journal} {J. Fluids Struct.}\ }\textbf
  {\bibinfo {volume} {2}},\ \bibinfo {pages} {355--381} (\bibinfo {year}
  {1988})}\BibitemShut {NoStop}%
\bibitem [{\citenamefont {Guckenheimer}\ and\ \citenamefont
  {Holmes}(2013)}]{guckenheimer2013}%
  \BibitemOpen
  \bibfield  {author} {\bibinfo {author} {\bibfnamefont {J.}~\bibnamefont
  {Guckenheimer}}\ and\ \bibinfo {author} {\bibfnamefont {P.}~\bibnamefont
  {Holmes}},\ }\href {https://books.google.com.ar/books?id=XYIpBAAAQBAJ} {\emph
  {\bibinfo {title} {Nonlinear Oscillations, Dynamical Systems, and
  Bifurcations of Vector Fields}}},\ Applied Mathematical Sciences\ (\bibinfo
  {publisher} {Springer New York},\ \bibinfo {year} {2013})\BibitemShut
  {NoStop}%
\bibitem [{\citenamefont {Govardhan}\ and\ \citenamefont
  {Williamson}(2000)}]{govardhan2000}%
  \BibitemOpen
  \bibfield  {author} {\bibinfo {author} {\bibfnamefont {R.}~\bibnamefont
  {Govardhan}}\ and\ \bibinfo {author} {\bibfnamefont {C.~H.~K.}\ \bibnamefont
  {Williamson}},\ }\bibfield  {title} {\enquote {\bibinfo {title} {{Modes of
  vortex formation and frequency response of a freely vibrating cylinder}},}\
  }\href@noop {} {\bibfield  {journal} {\bibinfo  {journal} {J. Fluid Mech.}\
  }\textbf {\bibinfo {volume} {420}},\ \bibinfo {pages} {85--130} (\bibinfo
  {year} {2000})}\BibitemShut {NoStop}%
\bibitem [{\citenamefont {Singh}\ and\ \citenamefont {Mittal}(2005)}]{Singh}%
  \BibitemOpen
  \bibfield  {author} {\bibinfo {author} {\bibfnamefont {S.}~\bibnamefont
  {Singh}}\ and\ \bibinfo {author} {\bibfnamefont {S.}~\bibnamefont {Mittal}},\
  }\bibfield  {title} {\enquote {\bibinfo {title} {Vortex-induced oscillations
  at low {R}eynolds numbers: {H}ysteresis and vortex-shedding modes.}}\
  }\href@noop {} {\bibfield  {journal} {\bibinfo  {journal} {Journal of Fluids
  and Structures.}\ ,\ \bibinfo {pages} {1085--1104}} (\bibinfo {year}
  {2005})},\ \bibinfo {note} {2005. 20(8): p}\BibitemShut {NoStop}%
\end{thebibliography}%

\end{document}